\documentclass{class/sig-alternate-05-2015}

\usepackage{tabularx}
\usepackage{booktabs}
\usepackage[square,comma,numbers,sort&compress,sectionbib]{natbib}
\usepackage{times}
\usepackage{url}
\usepackage{paralist}
\usepackage[usenames,dvipsnames]{color}
\usepackage{graphicx}
\usepackage{url}
\usepackage[bookmarks=true%
,bookmarksnumbered=true%
,hypertexnames=false%
,breaklinks=true%
,colorlinks=true%
,linkcolor=Blue%
,citecolor=Blue%
,urlcolor=Blue%
]{hyperref}
\usepackage{subfigure}
\usepackage{graphicx}
\usepackage{enumitem}
\usepackage{diagbox}
\usepackage{soul}
\usepackage{caption}
\usepackage{eurosym}
\usepackage{textcomp}
\usepackage{flushend}

\newcommand{\captionshrink}{}
\newcommand{\tablecaptionshrink}{}

\newcommand{\todo}[1]{\textcolor{red}{#1}}
\newcommand{\new}[1]{\textcolor{blue}{#1}}
\newcommand{\delete}[1]{\st{#1}}

\usepackage{algpseudocode,algorithm,algorithmicx}

\algnewcommand\algorithmicforeach{\textbf{for each}}
\algdef{S}[FOR]{ForEach}[1]{\algorithmicforeach\ #1\ \algorithmicdo}

\newcommand{\subs}[1]{$_{\text{#1}}$}

\begin{document}


%

\CopyrightYear{2017}
\setcopyright{acmlicensed}
\conferenceinfo{WSDM 2017,}{February 06 - 10, 2017, Cambridge, United Kingdom} 
\isbn{978-1-4503-4675-7/17/02}
\acmPrice{\$15.00} 
\doi{http://dx.doi.org/10.1145/3018661.3018679}

\clubpenalty=10000 
\widowpenalty = 10000

\title{Anticipating Information Needs Based on Check-in Activity}

\numberofauthors{3}

\author{}

 \author{
 \alignauthor
 Jan R. Benetka\\
    \affaddr{Norwegian University of Science and Technology}\\
    \email{benetka@idi.ntnu.no}
 \alignauthor
 Krisztian Balog\\
 	  \affaddr{University of Stavanger}\\
 	  \email{krisztian.balog@uis.no}
 \alignauthor
 Kjetil N{\o}rv{\aa}g\\
    \affaddr{Norwegian University of Science and Technology}\\
 	  \email{kjetil.norvag@idi.ntnu.no}
 }

\maketitle
\begin{abstract}

In this work we address the development of a smart personal assistant that is capable of anticipating a user's information needs based on a novel type of context: the person's activity inferred from her check-in records on a location-based social network.
Our main contribution is a method that translates a check-in activity into an information need, which is in turn addressed with an appropriate information card.  This task is challenging because of the large number of possible activities and related information needs, which need to be addressed in a mobile dashboard that is limited in size.
Our approach considers each possible activity that might follow after the last (and already finished) activity, and selects the top information cards such that they maximize the likelihood of satisfying the user's information needs for all possible future scenarios.  The proposed models also incorporate knowledge about the temporal dynamics of information needs.
Using a combination of historical check-in data and manual assessments collected via crowdsourcing, we show experimentally the effectiveness of our approach.


\end{abstract}


%
%

%
%



\section{Introduction}
\label{sec:introduction}

\begin{figure}[t]
    \centering
    \resizebox{\columnwidth}{!}{%
	\includegraphics[width=0.45\textwidth]{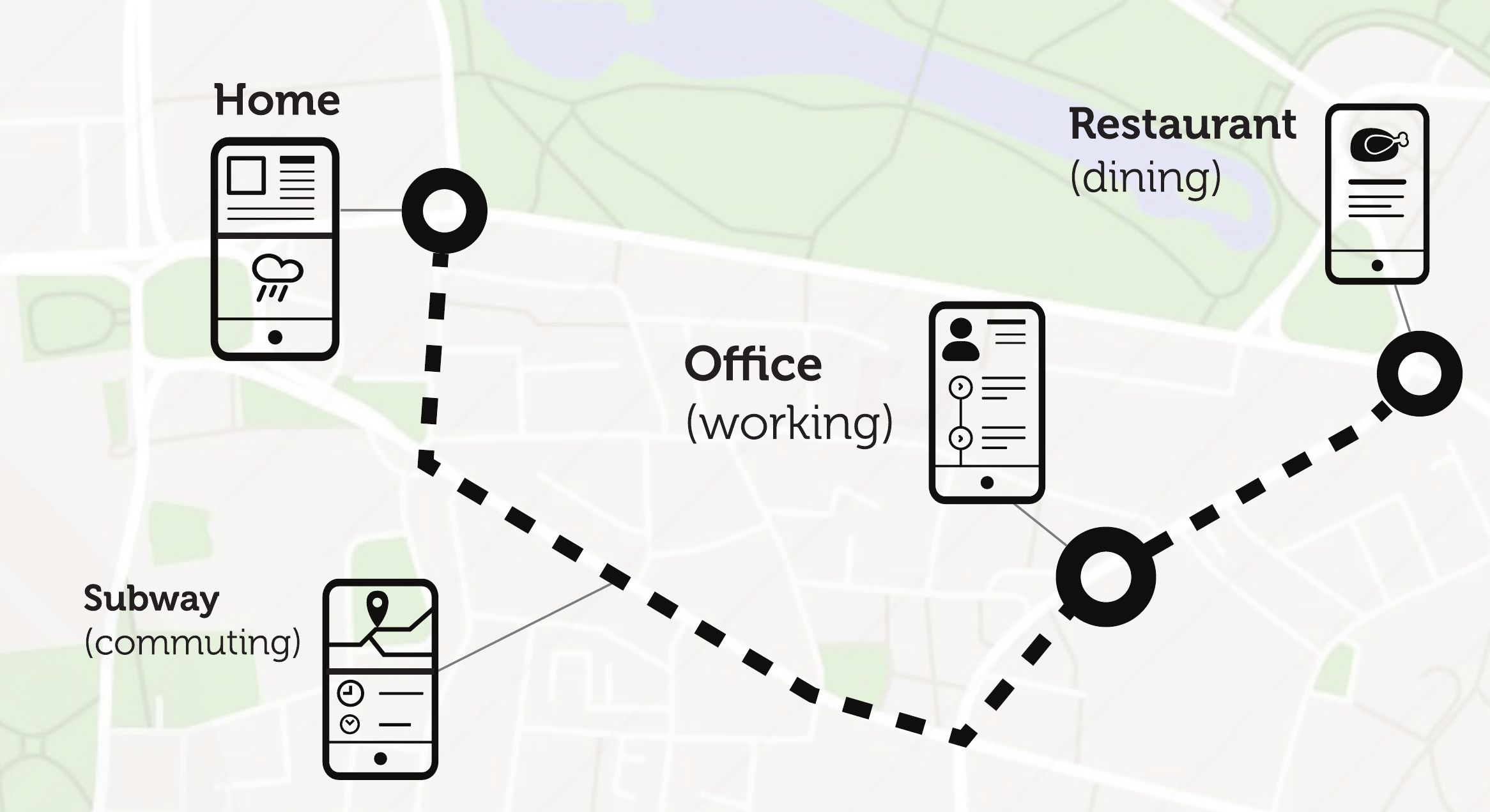}
	}
    \caption{Example information needs of a user during the course of a day, related to her current activity. A digital assistant should be able to anticipate these information needs, using the person's check-in activity as context, and proactively respond with a set of information cards that directly address those needs.}
    \label{fig:inf_need}
\end{figure}

Internet usage on mobile devices has been steadily growing and has now surpassed that of desktop computers. In 2015, Google announced that more than 50\% of all their searches happened on portable devices~\cite{mobile-searches}.
Mobile searches, to date, are still dominated by the conventional means, that is, using keyword queries~\cite{google-stats}.
Typing queries on a small device, however, is not necessarily comfortable nor is always easy.  Voice search and conversational user interfaces represent a promising alternative, by allowing the user to express her information need in spoken natural language~\cite{kamvar-2010-sww}.  Yet, this form of search may not be used in certain settings, not to mention that it will take some getting used to for some people to feel comfortable having their conversation with an AI in public.
Another main difference for mobile search is that it offers additional contextual information, such as current or predicted location, that can be utilized for improving search results~\cite{sun-2015-mul, krumm-2012-tam, amin-2009-fdc}.
Because the screens of mobile devices are rather limited in size, traditional list-based result presentation and interaction is not optimal~\cite{Church:2006:EII}.  A recent trend is to organize most useful pieces of information into \emph{information cards} \cite{Shokouhi:2015:QCR}; for example, for a restaurant, show a card with opening hours, menu, or current offers.
Importantly, irrespective of the means of querying, utilization of context, and presentation of results, these search systems still represent the traditional way of information access, which is \emph{reactive}.  A \emph{proactive} system, on the other hand, would anticipate and address the user's information need, without requiring the user to issue (type or speak) a query.  Hence, this paradigm is also known as \emph{zero-query search}, where ``systems must anticipate user needs and respond with information appropriate to the current context without the user having to enter a query''~\cite{allan-fco-2012}.  Our overall research objective is to develop a personal digital assistant that does exactly this: using the person's check-in activity as context, anticipate information needs, and respond with a set of information cards that directly address those needs.  This idea is illustrated in Figure~\ref{fig:inf_need}.
We tackle this complex problem by breaking it down into a number of simple steps.  Some of these steps can be fully automated, while others leverage human intelligence via crowdsourcing.

%

An \emph{activity}, in the scope of this paper, is defined as a category of a point-of-interest (POI) that the user visited, i.e., checked in to.  As argued in~\cite{Yang:2015:MUA}, a category is a very strong indicator of human activity. For instance, the category `\textit{football stadium}' implies watching or playing a football match.
We assume that this check-in information is made available to us, for instance, by means of a location-based social network application, such as the Foursquare mobile app. Alternatively, this information could be inferred to some extent from sensory data of a mobile device~\cite{partridge-2009-emr}.
The question for a proactive system then becomes: How to translate check-in activities to search queries?
Specifically, we consider a \emph{cold-start scenario}, in which we do not have access to mobile search query logs nor to behavioral data from past user interactions. 
Against this background, we ask the following two-fold research question:
\begin{itemize}
	\item[\textbf{RQ1}] How to identify common information needs and their relevance in the context of different activities? (\S\ref{sec:needs})
\end{itemize}
Using Foursquare's categories as our taxonomy of activities, we identify popular searches for each activity (i.e., POI category) by mining query suggestions from a web search engine for individual POIs (from the corresponding category).  
In a subsequent step, we then normalize search queries by grouping  together those that represent the same information need. As a result, we identify a total of $7,887$ distinct information needs for $287$ activities, which are organized in a two-level hierarchy. 
%
%



Presumably, different phases of an activity trigger different information needs.  To better understand the information requirements of people during various activities we ask the following question:
\begin{itemize}
	\item[\textbf{RQ2}] Do information needs change throughout the course of an activity (i.e., before, during, and after)? (\S\ref{sec:temporal})
\end{itemize}
Based on crowdsourced data, we show that the needs are dynamic in nature and change throughout the course of an activity; people ask for different types of cards before, during, or after an activity.  For example, before going to a nightlife spot, people welcome a card with information about dress code or opening hours, while during their visit a card with offers or menu is relevant. 

Having gained an understanding of information needs and their temporal nature, we turn to our ultimate task of anticipating a user's future information needs given her last activity.  We cast this task as a ranking problem:
\begin{itemize}
	\item[\textbf{RQ3}] How to rank future information needs given the last activity of the user as context? (\S\ref{sec:anticipating})
\end{itemize}
What makes this task challenging is that all possible future activities should be addressed on a single dashboard, which can display only a handful of cards. Thus, cards should be ranked in a way that they maximize the likelihood of satisfying the user's information need(s) for all possible future scenarios.
We introduce a number of increasingly complex probabilistic generative models that consider what activities are likely to follow next and what are the relevant information needs for each of those activities.
Evaluating the developed models is non-trivial; unless the envisioned proactive system is actually deployed, evaluation is bound to be artificial. To make the setup as close to a realistic setting as possible, we present a simulation algorithm that takes actual activity transitions from a large-scale check-in log.  We then collect manual judgments on information needs for a (frequency-based) sample of these transitions.
Our main finding is that models that address both (1) future information needs and (2) needs from the last activity are more effective than those that only consider (1). 
 

In summary, this paper makes to following novel contributions:
\begin{itemize}
\itemsep -2pt
    \item A method for obtaining information needs and determining their relevance for various activities without relying directly on a large-scale search log (\S \ref{sec:needs}). 
    \item A detailed analysis of how the relevance of information needs changes over the course of an activity for different categories of activities (\S\ref{sec:temporal}).
    \item A number of generative probabilistic models for ranking information needs given the user's last activity as context (\S\ref{sec:anticipating:model}).
    \item Evaluation methodology using a combination of a log-based simulator and crowdsourced manual judgments (\S\ref{sec:activity:expsetup}).  These evaluation resources are made publicly available.\footnote{\url{http://tiny.cc/zero-query-needs}}
\end{itemize}

\section{Related work}
\label{sec:related}

The idea of smart personal agents, which would help users to answer questions, find information or perform simple tasks has been around for at least a decade~\cite{mitchell-elp-1994,myers-ipa-2007}. Yet, it was only until recently that the advancement of technologies in AI, IR, and NLP, combined with a proliferation of mobile devices, allowed for wide-spread of these specialized applications. Commercial products such as Google Now~\cite{google-now}, Apple Siri~\cite{apple-siri}, and Microsoft Cortana~\cite{ms-cortana} are voice-controlled assistants built to automatically commit simple operational tasks on user's device or to search for information. Facebook~M~\cite{facebook-m} utilizes a hybrid approach which combines automated processing of information with human training and supervision. 
While the main focus is on processing explicit user commands, some of these systems are already capable of pre-fetching information based on users' behavioral patterns (e.g., Google Now). 

The concept of \emph{zero-query} (or \emph{proactive}) information retrieval has been first formalized at the Second Strategic Workshop on Information Retrieval~\cite{allan-fco-2012}, expressing the desire for systems that would anticipate information needs and address them without the user having to issue a query~\cite{liebling-2012-asu}, hence zero-query IR. Such systems are heavily dependent on user context, since it is the only source of information input. \citet{rhodes-2000-jitir} describe a just-in-time information retrieval agent which continuously monitors documents that a user is working with and presents related information without the user's intervention. The authors emphasize the importance of the agent's interface being non-intrusive and stress the priority of precision over recall. \citet{budzik-uie-2000} present a system that observes interactions with desktop applications, uses them to derive the user's textual context in highly routine situations, and proactively generates and issues a context-tailored query. \citet{braunhofer-2015-cam} propose a proactive recommender system that selects POIs to recommend based on user's preferences and pushes these suggestions only when the contextual conditions meet certain criteria (e.g., travel time to POI, weather). 
\citet{song-2016-qlp} take advantage of the repetitive nature of some tasks (e.g., reading news) to proactively suggest a next task to the user. This approach, however, is only applicable to a specific subset of tasks.

Query suggestion~\cite{Liao:2011:MCS,Kato:2013:PUQ} and auto-completion~\cite{Bar-Yossef:2011:CQA,Shokouhi:2013:LPQ} are fundamental services in modern search engines.
The general idea behind them is to assist users in their search activities by \emph{anticipating} their information needs; the provided suggestions help users to articulate better search queries.  The underlying algorithms draw on query reformulation behavior of many searchers, as observed in large-scale search logs~\cite{White:2016:ISS}.
Recently, a special emphasis has been placed on recommending queries that aid users in completing complex (multi-step or multi-aspect) search tasks~\cite{HassanAwadallah:2014:SCS,Yilmaz:2015:OTT}.
Importantly, all these suggestions are for refining an existing (or incomplete) query, which is unavailable in zero-query IR.  (Instead, the user's context may serve as the initial query.)  

\citet{sohn-2008-ads} report that $72\%$ of mobile information needs are triggered by one of the following contexts: activity, location, time, or conversation.  \citet{hinze-2010-cqe} find that $50\%$ of mobile needs are influenced by location and activity and $16\%$ alone by activity. In this paper, we consider activities, represented by POI categories, as context.  POI categories have been used in prior work for activity prediction~\cite{noulas-esg-2011, Yang:2015:MUA}.
Categories have also been exploited in POI-based recommendation to reduce the pool of candidate venues \cite{sang-asc-2015, liu-ppr-2013, zhang-geosoca-2015}. \citet{kiseleva-2013-pcu} use categories to find sequences of users' activities in a web-browsing scenario. They extend Markov models with geographical context on a continent level. Another approach using (personalized) Markov chains is introduced in~\cite{cheng-wyl-2013}. The authors address the task of successive POI recommendation while imposing geographical constraints to limit the number of possible POI candidates. Similar techniques could be exploited for the next activity prediction, a subtask in our approach (cf. \S\ref{sec:anticipating:pattern}). 

It is a recent trend to address information needs in the form of domain-specific information cards. \citet{Shokouhi:2015:QCR} discover temporal and spatial (i.e., work/home) implications on user interactions with the cards and propose a card ranking algorithm called Carr\'e. In~\cite{guha-2015-ump, yang-2016-mui}, the authors focus on modeling user interests to better target user needs within personal assistants. In both cases a commercial query log is used as a source of data. \citet{hong-2016-loc} study the ranking of information cards in a reactive scenario, i.e., with user-issued queries. They propose an approach for interpreting query reformulations as relevance labels for query-card pairs, which in turn are used for training card ranking models.

\section{Information needs related to \\activities}
\label{sec:needs}

In this section, we define activities (\S\ref{sec:needs:activities}),  present a semi-automatic approach for identifying and ranking information needs with respect to their relevance given an activity  (\S\ref{sec:needs:collect}), and evaluate the proposed method in a series of crowdsourcing experiments (\S\ref{sec:needs:eval}).

\subsection{Activities}
\label{sec:needs:activities}

We define an \emph{activity} as the category of a point-of-interest (POI) that the user visited, i.e., checked in to.  In the remainder of this paper, we will use the terms activity and category interchangeably.\footnote{Admittedly, a given POI category might imply a set of different activities.  For example, visitors at a beach could bathe, jog, stroll on the promenade, or relax at a caf\'{e}.  Nevertheless, the category is a good indicator of the scope of possible pursuits; requiring users to provide more detailed account of their activities upon checking in to a POI would be unreasonable in our opinion.} 
Activities may be organized hierarchically in a taxonomy.  When the hierarchical level is indifferent, we simply write $a \in A$ to denote an activity, where $A$ is the universe of all activities; otherwise, we indicate in the superscript the hierarchical level of the activity, i.e., $a^{l1}$ for top-level, $a^{l2}$ for second level, and so on.  
We base our activity hierarchy on Foursquare, as further detailed below.  We note that this choice is a rather pragmatic one, motivated by the availability of data.  The approaches presented in this paper are generic and could be applied to arbitrary activities given that a sufficient number of POIs is available for each activity.  

\subsubsection{Check-in data} 
\label{sec:needs:activities:fsdata}


Foursquare is a location-based social network heavily accessed and used via a mobile application for local search and discovery. Registered users check-in to POIs, which are organized into a 3-tier hierarchy of POI categories with $10$ top-level, $438$ second-level and $267$ third-level categories.\footnote{\url{https://developer.foursquare.com/categorytree}} 
We make use of the TIST2015 dataset~\cite{yang-2016-pcm},\footnote{\url{\detokenize{http://bit.ly/datasets-dingqi_yang}}} which contains long-term check-in data from Foursquare collected over a period of 18 months (Apr 2012--Sept 2013).  It comprises $33$M check-ins by $266.9$K users to $3.68$M locations (in $415$ cities in $77$ countries). 
Each POI in TIST2015 is assigned to one of the Foursquare categories from an arbitrary level, with the majority ($84$\%) of POIs assigned to a second-level category. 

We create our activity hierarchy by taking the top two levels of Foursquare's POI categories and populate them with POIs from the TIST2015 dataset.  For each second-level category, we take the top $200$ most visited POIs as a representative sample of that category.  Further, we limit ourselves to POIs from English speaking countries.\footnote{Australia, United Kingdom, Ireland, New Zealand, USA and South Africa} 
We keep only non-empty categories (i.e., that contain POIs that meet the above requirements).  As a result, we end up with $9$ top-level and $287$ second-level categories. 
Since the dataset does not contain the names of POIs, we use the Foursquare API to obtain the names of the sampled POIs. 

%
%

\subsection{Method}
\label{sec:needs:collect}

Our objective is to identify information needs and establish their relevance for a given activity.  Formally, we need to obtain a set of information needs, $I$, and estimate the probability $P(i|a)$ of each information need $i \in I$, which expresses its relevance with respect to a given activity $a \in A$.  

This is a non-trivial task, especially in a cold-start scenario, when no usage data had been generated that could be used for establishing and further improving the estimation of relevance. 
It is reasonable to assume that common activity-related information needs are reflected in the search queries that people issue~\cite{hinze-2010-cqe}. In order to make our method applicable in cold-start scenario (and outside the walls of a major search engine company), we opt not to rely directly on a large-scale search log.
We attempt to gain indirect access by making use of search query completions provided by web search suggestions.  By analyzing common search queries that mention specific instances of a given activity, we can extract the most frequent information needs related to that activity. 
Below, we present the technical details of this process and the normalization steps we applied to group search queries together that represent the same information need.


\subsubsection{Collecting query suggestions}



For each second-level POI category, we take all sampled POIs from that category (cf.~\S\ref{sec:needs:activities:fsdata}) and use them as ``query probes.''  This process resembles the query-based collection sampling strategy used in uncooperative federated search environments~\cite{Shokouhi:2011:FS}.

\begin{figure}[h]
    \centering
	\includegraphics[width=0.48\textwidth]{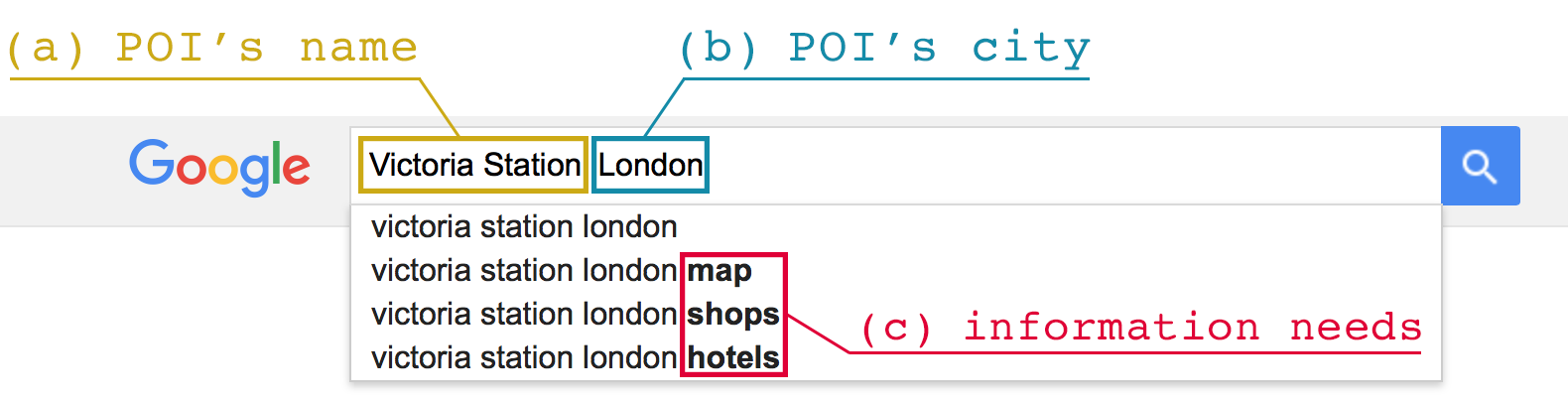}
    \caption{Query suggestions by Google; each query is a combination of POI's name (a) and location (b). The completions (c) represent information needs related to the POI.}
    \label{fig:gs}
\end{figure}

The query is created as a concatenation of the POI's name and location (city), and used as input to the Google Query Suggestion API.\footnote{\url{https://www.google.com/support/enterprise/static/gsa/docs/admin/70/gsa_doc_set/xml_reference/query_suggestion.html}} This API returns a list of (up to $10$) top-ranked suggestions as a result; see Figure~\ref{fig:gs}.  

\noindent
The list includes the original query, the most popular completions (searches with the same prefix), as well as possible reformulations.  We ignore the reformulations and extract suggested suffixes (e.g., `map,' `opening hours') as individual information needs, which we then aggregate on the category level. It should be noted that the suggestions are not personalized, since the API calls do not contain any explicit information about the user.




We collected suggestions for $287$ second-level POI categories, using the top (up to) $200$ POIs from each, resulting in a total of $44,342$ POIs for generating queries; 73\% of these led to a non-empty set of suggestions. We obtained a total of $83,658$ suffixes. Before further processing, the following cleansing steps were applied: (1) removal of numbers (`fashion week \delete{2016}'), day and month names (`opening hours \delete{january}'), and geographical references (e.g., `\delete{ohio} store hours'); and (2) restriction to terms longer than $2$ characters.
At the end of this process we are left with a total of $66,177$ suggestion suffixes that are aggregated on the category level. Table~\ref{tab:checkin_distrib} displays the distribution of suffixes across the top-level categories.

\begin{table}[t]
    \begin{center}
    \caption{Query suggestions, after data cleaning, aggregated per POI category and ordered by frequency.}
    \tablecaptionshrink
    \label{tab:checkin_distrib}
    \begin{tabular}{ l  r  r } 
\toprule[1.2pt]
\textbf{Category name} & \multicolumn{2}{c}{\textbf{\#suggestions}} \\ 
& \textbf{total} & \textbf{unique} \\ 
\midrule
Food  							& $25072$ 	& $3080$	\\ 
Shop \& Service 				& $18505	$	& $3293$	\\ 
Arts \& Entertainment 			& $5747$ 	& $1268$	\\ 
Outdoors \& Recreation 			& $4871$ 	& $1172$	\\ 
Nightlife Spot 					& $4898$ 	& $957$	\\ 
Professional \& Other Places 	& $2877$ 	& $926$	\\ 
Travel \& Transport 			& $2148$ 	& $721$	\\ 						        
College \& University 			& $1873$ 	& $617$ 	\\ 
Residence 						& $186$ 		& $85$		\\ 
\bottomrule[1.2pt]
    \end{tabular}
    \end{center}
\end{table}


\subsubsection{Normalization} 
\label{sec:needs:norm}

Information needs, as obtained from the query suggestions, are typically expressed in a variety of ways, even when they have the same meaning.  We will simply refer to these ``raw'' information needs as \emph{terms}, noting that they may actually be phrases (i.e., multiple terms).  We wish to normalize the collected suggestions into a canonical set, such that all terms that express the same underlying information need are grouped together; see Table~\ref{tab:in_synonyms} for examples. We took the $100$ most frequent terms from each category and let three assessors, including the first author of this paper, group synonyms together.  The inter-assessor agreement as measured by Fleiss' kappa was $\kappa=0.52$, which is considered to be moderate agreement.  Each assessor $x_i \in X$ created $m_i$ sets of synonyms $s_1,\dots,s_{m_i} \in S_{x_i}$ from the extracted terms $T$.  In order to merge the collected results, while keeping the logical separation of synonyms, we use a graph-based approach.  We build an undirected graph $G$ where nodes $N$ correspond to terms $T$ and edges $E$ connect the terms that belong to the same synonym set $s_m$. 
In this graph, the terms that are grouped together by multiple assessors will form densely connected areas of nodes.  To separate these areas we use the DPClus graph clustering algorithm~\cite{price:2013:survey}.  Finally, we label each cluster manually with a canonical name, which is typically the most frequent term within the cluster. 
In total, after normalization we recognize $7,887$ distinct information needs.


%
%

\begin{table}[ht!]
    \begin{center}
    \caption{Information need labels and their synonym terms.}
    \tablecaptionshrink
    \label{tab:in_synonyms}
    \begin{tabular}{ l  l } 

\toprule[1.2pt]
\textbf{Information need} & \textbf{Synonyms} \\
\midrule
jobs 			& employment, job, careers, career, \dots \\ 
map  			& localization map, map, travel maps, \dots \\ 
prices 			&  price list, price, prices, costs, taxi rate, \dots \\ 
operation hours & opening time, office hours, times, \dots \\ 
\toprule[1.2pt]

    \end{tabular}
    \end{center}
    \captionshrink
\end{table}

Table~\ref{tab:in_synonyms} lists some information needs and their synonyms; the term `operation hours' for example has 61 synonyms in our dataset. 
In the remainder of the paper, when we talk about information needs, we always mean the normalized set of information needs.

\subsubsection{Determining relevance}
\label{sec:needs:relevance}
The ranking of the extracted and normalized information needs is defined by their relative frequency, because, intuitively, the more often people search for a query, the more relevant the information need it represents.  Formally, let $n(i,a)$ denote the number of times information need $i$ appears for activity $a$. We then set
\begin{equation}
	P(i|a) = \frac{n(i,a)}{\sum_{i' \in I}n(i',a)},	
	\label{eq:pia}
\end{equation}
where $I$ is the set of distinct information needs.

\subsubsection{Analysis}
\begin{figure*}[ht!]
    \centering
	\includegraphics[width=1\textwidth]{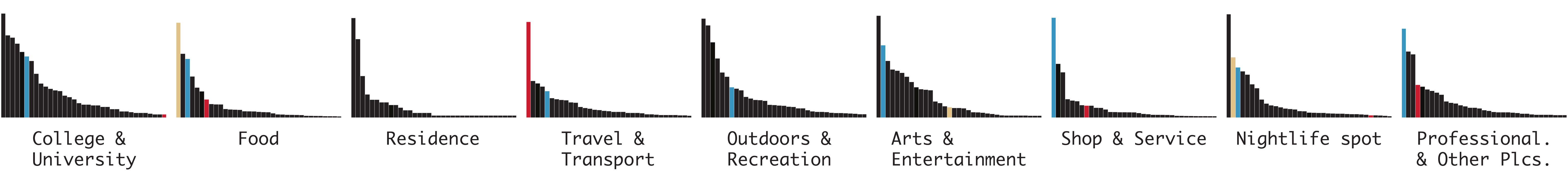}
    \caption{Distributions of information needs per category (top 35 depicted). Bars represent information needs, the size of the bars is proportional to the number of times the information need appears for that activity ($n(i,a)$). Highlighted information needs are `operating hours' (blue), `menu' (yellow), and `airport' (red).}
    \label{fig:cat_distributions}
    \captionshrink
\end{figure*}

Information needs follow a tail-heavy distribution in each top-level category; the head information needs are shown in Figure~\ref{fig:cat_distributions}.  On average, the top 25 information needs in each category cover $59$\% and $72$\% of all information needs for top- and second-level categories, respectively. 
Not surprisingly, some categories have a larger portion of domain-specific information needs, such as the `College \& University' category with terms like \emph{`university info,' `campus,'} or \emph{`study programme.'}  On the other hand, some information needs are almost universally relevant: \emph{`address,' `parking,'} or \emph{`operation hours.'} 
To measure how (dis)similar information needs are across categories, we compute the Jaccard coefficient between the top $10$ information needs of each category, for all pairs of top-level categories. We find that the categories are very dissimilar in terms of information needs on the top positions. The closest are `Nightlife spot' and `Food,' with a similarity score of $0.3$.

%

%
%
%

\subsection{Evaluation}
\label{sec:needs:eval}

Next, we evaluate the performance of our method by measuring the recall of the extracted information needs (\S\ref{sec:needs:eval:recall}) and their ranking w.r.t. relevance (\S\ref{sec:needs:eval:rank}).
For both, we compare the extracted information needs against crowdsourced human judgments.\footnote{Details of the crowdsourcing experiments are delegated to an online appendix \url{http://tiny.cc/zero-query-needs}.}

\subsubsection{Evaluating recall}
\label{sec:needs:eval:recall}

In the first crowdsourcing experiment, we seek to measure the recall of the extracted information needs.  We ask people to imagine being at a location from a given top-level POI category and provide us with the top three information needs that they would search for on a mobile device in that situation.  

Due to its free-text nature, manual inspection and normalization had to be applied to the user input.  $51$ entries had to be removed due to violation of rules, such as inputting text in different language or duplicating the same entry for all three fields.  In total, we received $712$ valid answers.  We mapped information needs that have previously been identified to the corresponding normalized version (cf.~\S\ref{sec:needs:norm}); otherwise, we treated it as a unique information need.  Note that this is a pessimistic scenario, assuming that all these unseen information needs are distinct.  It may be that some of them could be clustered together, therefore, the evaluation results we present should be regarded as lower bounds. Another factor negatively influencing the recall values is the limitation of the human-based normalization process, in which only the most frequent terms are considered for each category (cf.~\S\ref{sec:needs:norm}). For instance, in the `Nightlife Spot' category, the information need \emph{`party'} is not recognized, even though terms like \emph{`Christmas party,'} \emph{`private party,'} or \emph{`foam party'} exist in the long tail of the suggestions distribution.
Table~\ref{tab:recall} presents the results at different recall levels.
We observe very similar Recall@10 for all categories except of `Food', which stands out. This category also exhibits very high Precision@10 of $0.8$. Particularly low recall ($0.28$@All) is obtained for the `Residence' category, which may be caused by the fact that POIs within this category are in many cases homes of users and therefore generate only a few suggestions.




\begin{table}[ht!]
    \begin{center}
    \caption{Evaluation of recall at various cutoff points. \#Needs is the number of norm. information needs according to the ground truth.} 
    \label{tab:recall}
    \tablecaptionshrink
    \resizebox{\columnwidth}{!}{%
    \begin{tabular}{ l@{~~}r@{~~}r@{~~}r@{~~}r } 

\toprule[1.2pt]
\textbf{Category} 			& \textbf{\#Needs} & \textbf{R@10} 	& \textbf{R@20} 	& \textbf{R@All} \\
\midrule


College \& University 		 & 27 & 0.22 & 0.37 & 0.74 \\ 
Food  						 & 15 & 0.53 & 0.53 & 0.73 \\ 
Residence 					 & 36 & 0.22 & 0.25 & 0.28 \\ 
Travel \& Transport 		 & 25 & 0.24 & 0.36 & 0.48 \\ 
Outdoors \& Recreation 		 & 19 & 0.26 & 0.53 & 0.89 \\ 
Arts \& Entertainment 		 & 22 & 0.23 & 0.27 & 0.68 \\ 
Shop \& Service 			 & 22 & 0.23 & 0.36 & 0.77 \\ 
Nightlife Spot 				 & 18 & 0.33 & 0.50 & 0.78 \\ 
Professional \& Other Places & 31 & 0.26 & 0.39 & 0.65 \\ 

\midrule 
Average						& 23.9 & 0.28 & 0.40 & 0.67 \\ 
\bottomrule[1.2pt]

    \end{tabular}
    }
    \end{center}
    \captionshrink
\end{table}

\subsubsection{Evaluating relevance}
\label{sec:needs:eval:rank}

\begin{figure}[t]
    \centering
	\fcolorbox{Gray}{white}{\includegraphics[width=0.46\textwidth]{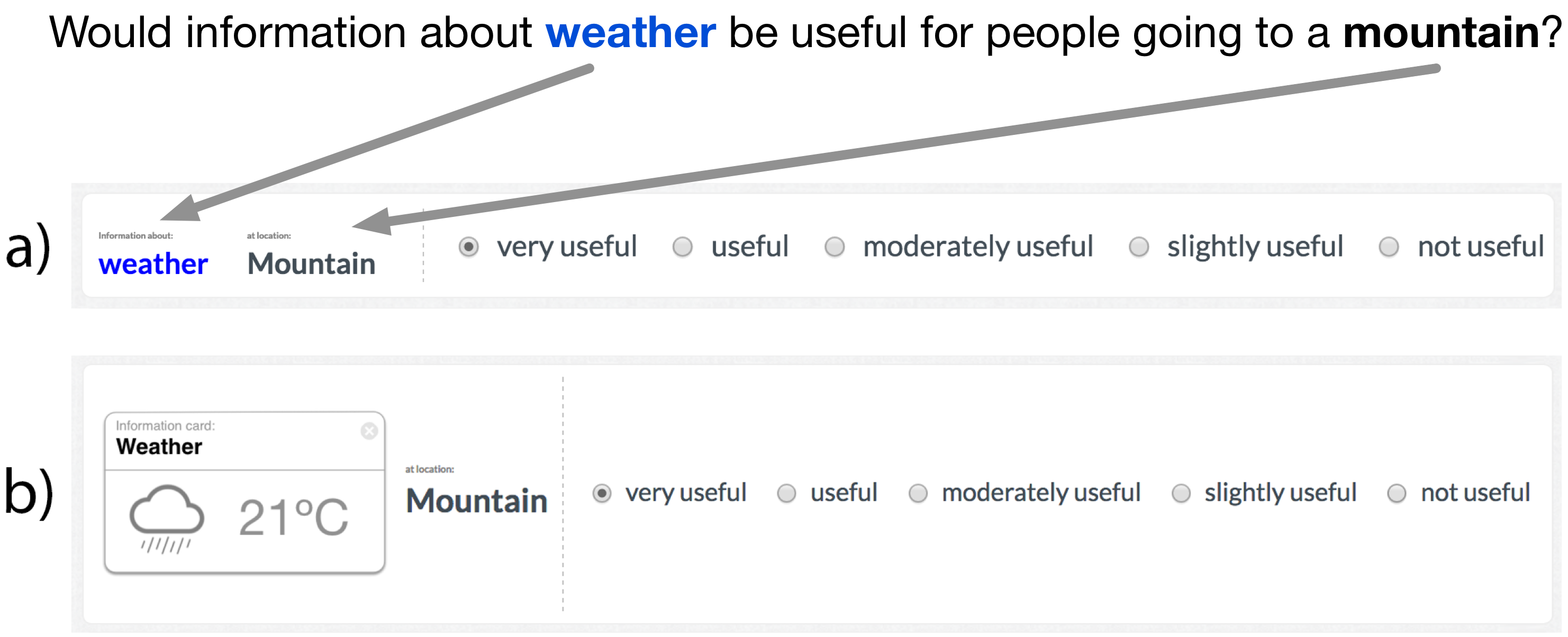}}
    \caption{Crowdsourcing experiment \#2, asking to rate the usefulness of a given information on the second-level POI category in a) textual form and b) card-based form.}
    \label{fig:csexp2}
\end{figure}

Our second set of experiments is aimed at determining how well we can rank information needs with respect to their relevance given an activity (i.e., $P(i|a)$).
We conduct two experiments: first in textual mode and then in more visually oriented card-based form; see Figure~\ref{fig:csexp2} top vs. bottom.
This comparison allows us to examine if the presentation form changes in any way the perception and valuation of the actual content.

In both cases, we ask study participants to rank the usefulness of a given information need with respect to a selected category on a 5-point Likert scale, from \emph{`not useful'} to a \emph{`very useful'} piece of information. 
We evaluated the top $25$ information needs for the $5$ most visited second-level categories for each of the $9$ top-level categories, amounting to $25 \times 5 \times 9=1125$ distinct information need and activity pairs. 
We computed the Pearson's correlation for the two variants, i.e., textual and card-based, and found strong correlation: $0.91$ and $0.77$ for the top and second-level activities, respectively. Crowdsourcing workers' satisfaction was slightly higher in the card-based variant, which indicates that visual input is easier to grasp than plain text.

Table~\ref{tab:ranking} presents the evaluation results in terms of NDCG. We find that both variants achieve comparable results with card-based method performing better at the cutoff position of $3$ and being worse at the rest of measured positions. The differences, however, are negligible.

\begin{table}[h!]
    \begin{center}
    \caption{Evaluation of the ranking of information needs with respect to their relevance for a given activity.}
    \label{tab:ranking}
    \tablecaptionshrink
    \resizebox{\columnwidth}{!}{%
    \begin{tabular}{ l r r r} 

\toprule[1.2pt]
\textbf{Ground truth} & \textbf{NDCG@3} & \textbf{NDCG@5} & \textbf{NDCG@10} \\
\midrule
Text-based & 0.491 & 0.550 & 0.627 \\
Card-based & 0.519 & 0.535 & 0.603 \\
\bottomrule[1.2pt]

    \end{tabular}
    }
    \end{center}
    \captionshrink
\end{table}

\section{Analysis of Temporal Dynamics of Information Needs}
\label{sec:temporal}

In this section we test our hypothesis that the relevance of an information need may vary during the course of an activity (cf. RQ2).

\subsection{Method}
\label{sec:temporal:method}

We define the following three temporal periods ($t$) for an activity:

\begin{itemize}
\itemsep -1pt
	\item{\textbf{Period before an activity (`pre')}} -- information is relevant before the user starts the activity; after that, this information is not (very) useful anymore.
	\item{\textbf{Period during an activity (`peri')}} -- information is mainly relevant and useful during the activity.
	\item{\textbf{Period after an activity (`post')}} -- information is still relevant to the user even after the actual activity has terminated.
\end{itemize} 
We introduce the concept of \emph{temporal scope}, which is defined as the probability of an information need being relevant for a given activity during a certain period in time.  
In the lack of a mobile search log (or similar resource), we resort to crowdsourcing to estimate this probability:
\begin{equation}
	\label{eq:ts}
	P(t|i,a) = \frac{n(t,i,a)}{\sum_{t \in \{pre,peri,post\}} n(t,i,a)},
\end{equation}
where $n(t,i,a)$ is the number of votes assigned by crowdsourcing workers to the given temporal period $t$ for an information need $i$ in the context of an activity $a$. 




\subsection{Experimental setup}

\begin{figure}[t]
    \centering
    \fcolorbox{Gray}{white}{\includegraphics[width=0.46\textwidth]{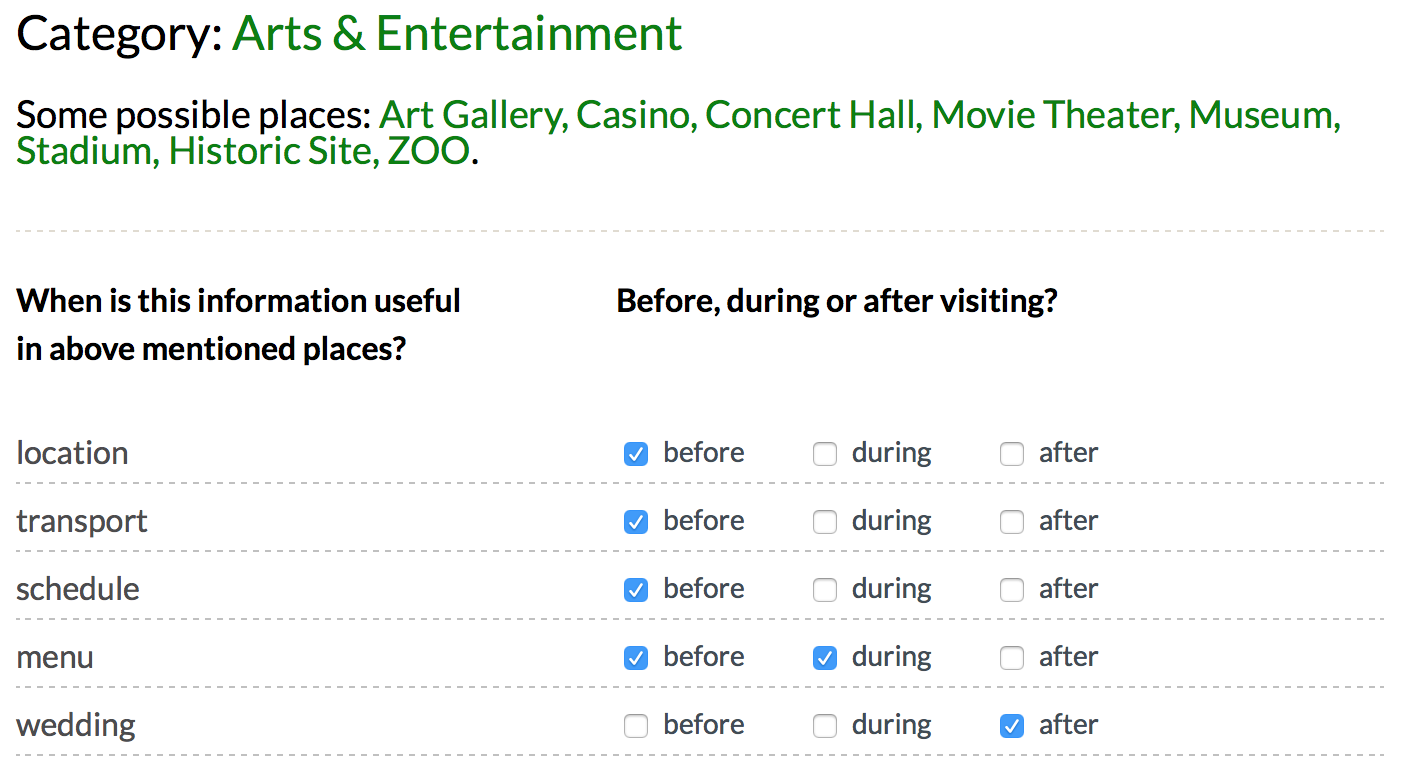}}
    \caption{Crowdsourcing experiment \#3, requiring users to specify the time period when a given information is the most useful with respect to a certain activity.}
    \label{fig:csexp3}
\end{figure}

We set up the following crowdsourcing experiment to collect measurements for temporal scope.  In batches of $5$, we presented the $30$ top-ranked information needs in each top-level category.  The task for the assessors was to decide when they would search for that piece of information in the given activity context: before, during, or after they have performed that activity.  They were allowed to select one or more answers if the particular information need was regarded as useful for multiple time slots. Figure~\ref{fig:csexp3} depicts the assessment interface.
In order to validate the collected data, we ran this experiment twice and compared data from both rounds.  In the first run, we had each information need processed by at least $5$ workers and in the second run we required at least $4$ more.  A cosine similarity of $85$\% suggests that participants were consistent in judging the temporal scope of individual information needs in the two experimental runs.


\subsection{Results and analysis}
\begin{figure}[t]
    \centering
	\includegraphics[width=0.48\textwidth]{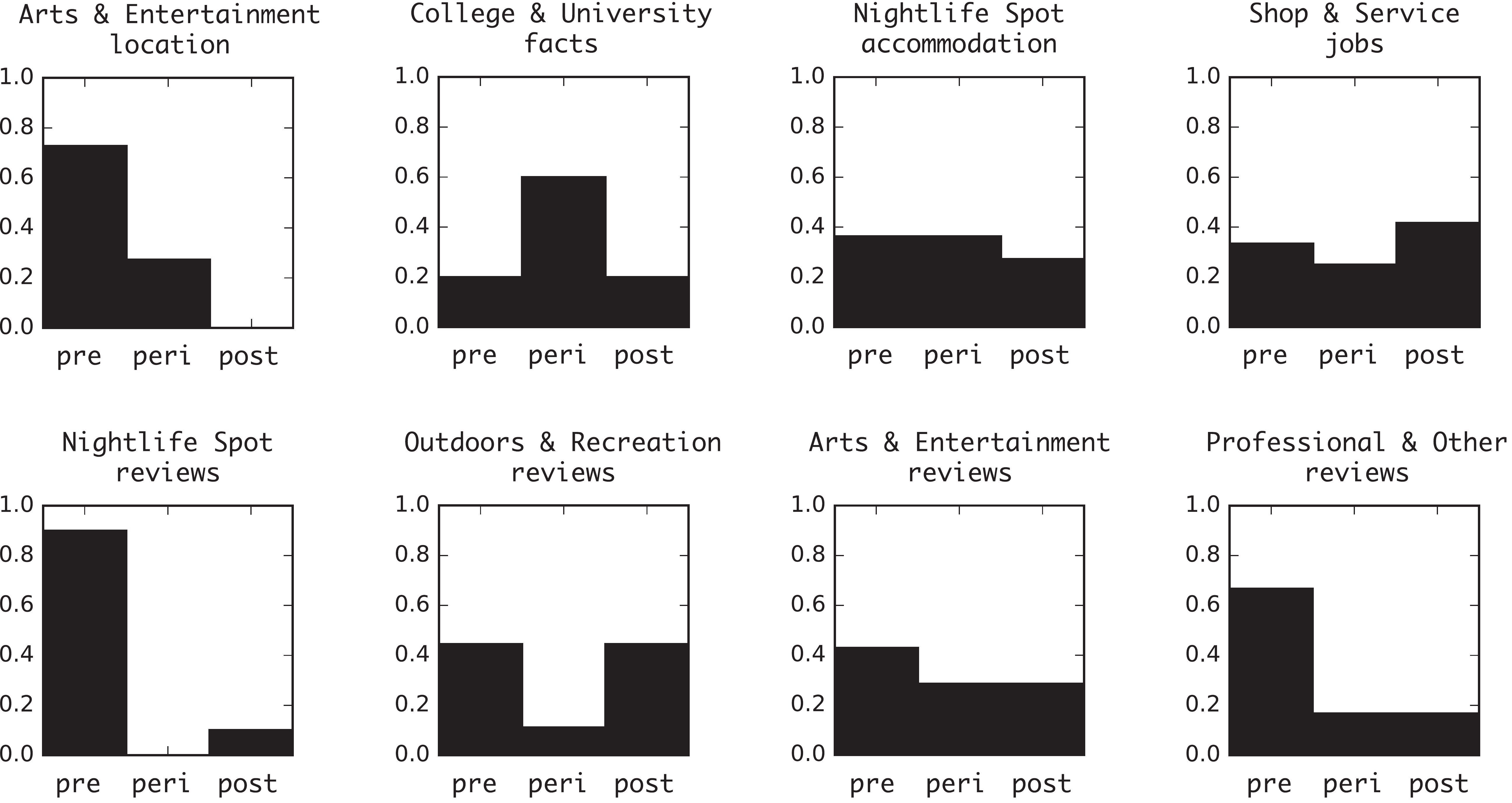}
    \caption{Distribution of temporal scopes ($P(t|i,a)$) for a selection of activity and information need pairs. Notice that the figures in the bottom row all belong to the same information need (\emph{`reviews'}), but the activities are different.}
    \label{fig:temp_scope}
    \captionshrink
\end{figure}

Figure~\ref{fig:temp_scope} plots temporal scopes for a selection of information needs and activities.  We can observe very different temporal patterns, confirming our intuition that information needs do change throughout the course of an activity.

\begin{figure}[t]
    \centering
	\includegraphics[width=0.46\textwidth]{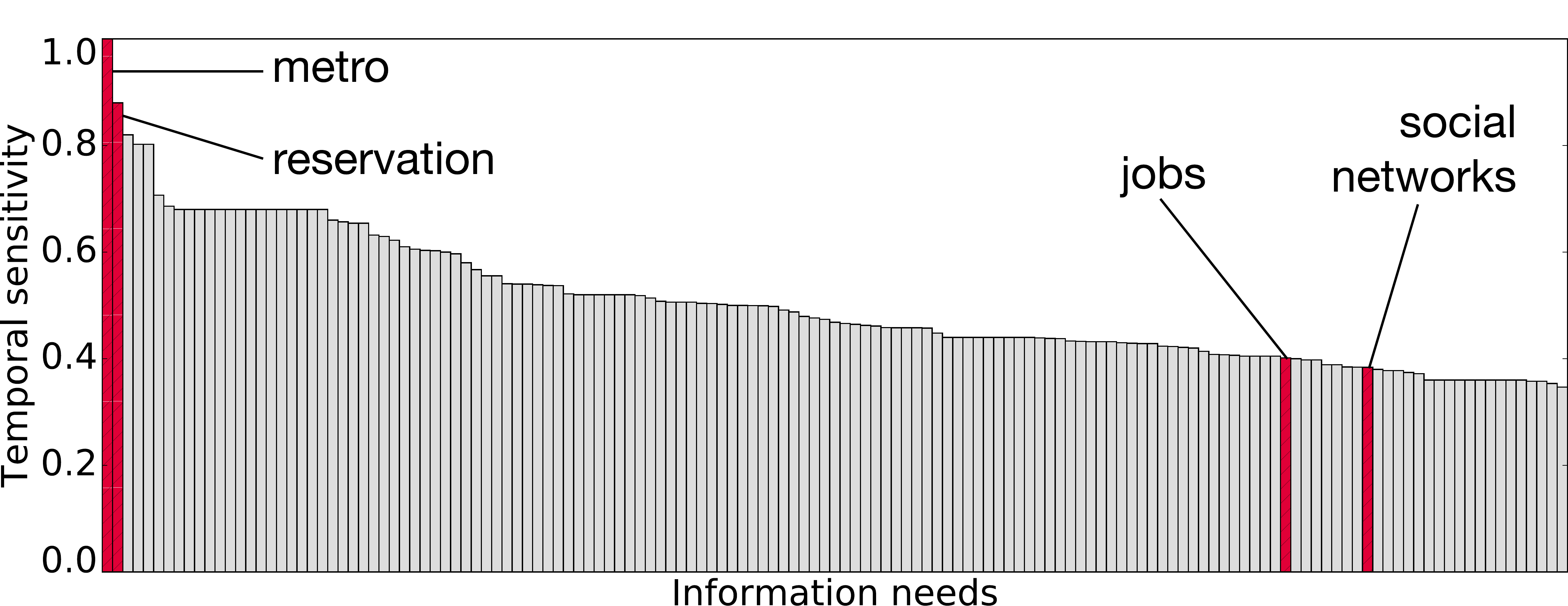}
    \caption{Temporal sensitivity of information needs.}
    \label{fig:temp_sens_needs}
    \captionshrink
\end{figure}

Further, we introduce the notion of \emph{temporal sensitivity} (TS), to characterize the dispersion of the information need's temporal scope.  We define it as the variance of temporal scope:
\begin{equation}
	TS(i,a) = \mathrm{Var}(P(.|i,a)).
\end{equation}
Temporal sensitivity reflects how salient is the right timing of that particular information need for a given activity. 
Figure~\ref{fig:temp_sens_needs} displays TS of information needs (averaged if belongs to multiple categories). 

\section{Anticipating Information Needs}
\label{sec:anticipating}

\begin{figure}[ht!]
	\includegraphics[width=0.48\textwidth]{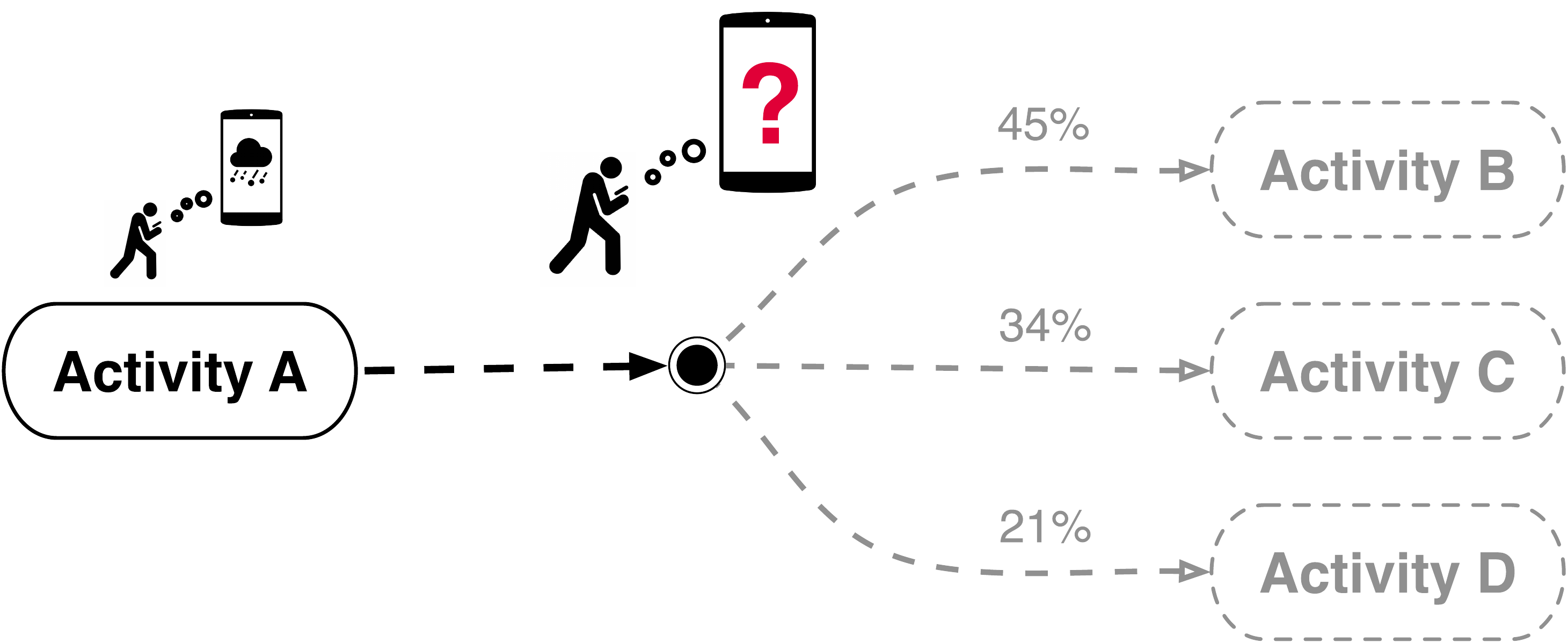}
    \caption{Anticipating a user's information needs after a given activity (\textit{Activity A}).}
    \label{fig:model}
\end{figure}

So far, we have identified information needs related to a given activity (\S\ref{sec:needs}) and studied how their relevance changes over the course of the activity (\S\ref{sec:temporal}).  We have shown that some information needs are important to address before the actual activity takes place and for some other needs the reach lasts even after the activity has terminated. 
Recall that our goal is to develop a \emph{proactive} mobile application. We assume that this system has information about the last activity of the user ($a_{last}$), that is, the category of the last check-in.  
The system shall then \emph{anticipate} what information need(s) the user will have next and address these needs by showing the corresponding information cards on the mobile dashboard proactively.  To be able to do that, the system needs to consider each possible activity that might follow next ($a_{next}$) and the probability of that happening ($P(a_{next}|a_{last})$).  Then, the top information needs to be shown on the dashboard are selected such that they maximize the likelihood of satisfying the user's information need(s) for all possible future scenarios.  This idea is depicted in Figure~\ref{fig:model}.

\noindent
We note that determining the exact timing for displaying information cards proactively is an interesting research question; however, we leave this to future work. Our focus in this work is on determining \emph{what} to show to the user and not on \emph{when} to show it. What is important for us is that the information need gets addressed \emph{before} the user embarks on her next activity.


\subsection{Models}
\label{sec:anticipating:model}


We formulate our problem as the task of estimating the probability (of relevance) of an information need given the last activity, $P(i|a_{last})$. This probability is computed for all information needs ($I$), then the top-$k$ ones are selected to be addressed (by displaying the corresponding information cards on the mobile dashboard).

In the following, we introduce three increasingly complex models.  These share the same components (even though not all models may not utilize all of these components):
\begin{itemize}
\itemsep -1pt
	\item $P(i|a)$: the relevance of an information need given an activity.  We established this quantity in~\S\ref{sec:needs:activities}, cf. Eq.~\eqref{eq:pia}, but we discuss further refinements in \S\ref{sec:anticipating:model:pia}.  Note that his probability is not to be confused with $P(i|a_{last})$ that we wish to establish.  
	\item $P(t|i,a)$: the temporal scope of an information need for a given activity.  This probability is estimated based on manual judgments, cf.~\S\ref{sec:temporal:method} and Eq.~\eqref{eq:ts}; see~\S\ref{sec:anticipating:model:temp} for further details.
	\item $P(a_{next}|a_{last})$: transition probability, i.e., the likelihood of activity $a_{last}$ being followed by $a_{next}$.  We introduce a method for estimating this probability from a check-in log data in~\S\ref{sec:anticipating:pattern}.
\end{itemize}

\noindent
Our first model, \textbf{M1}, considers all possible upcoming activities:
\begin{equation}
   P_{M1}(i|a_{last}) = \sum_{a_{next}} P(i|a_{next})P(a_{next}|a_{last}).  \nonumber
\end{equation}
\noindent
The second model, \textbf{M2}, is a linear mixture of two components, corresponding to the probability of the information need given (1) the last activity and (2) the upcoming activity.  The interpolation parameter $\gamma \in [0,1]$ expresses the influence of the last activity, i.e., the extent to which we want to address post information needs of the last activity. Formally,
\begin{equation*}
\begin{aligned}
   P_{M2}(i|a_{last}) = & \gamma P(i|a_{last}) \\
   & + (1-\gamma)\sum_{a_{next}} P(i|a_{next})P(a_{next}|a_{last}). \\ 	
\end{aligned}
\end{equation*}
We set the parameter $\gamma$ according to the average post-relevance of information needs across all activities:
\begin{equation}
   \gamma = \frac{\sum_{i \in I,a \in A} P(t_{post}|i,a)}{|I|\cdot |A|}, 
   \label{eq:gamma}
\end{equation}
where $P(t_{post}|i,a)$ is the post-relevance of the information need, $|A|$ is the number of all activities on the top-level, and $|I|$ is the cardinality of the set of all possible information needs.

Notice that according to this second model, the post-relevance of the last activity is the same for all activities and information needs.  Clearly, this is a simplification.  
Our final model, \textbf{M3}, is further extension that considers the temporal dynamics of each information need individually:
\begin{equation*}
\begin{aligned}	
   P_{M3}(i|a_{last}) & \propto P(t_{post}|i,a_{last}) P(i|a_{last}) \\
   & + \sum_{a_{next}} P(t_{pre}|i,a_{next})P(i|a_{next})P(a_{next}|a_{last}).
\end{aligned}
\end{equation*}
Specifically, $\gamma$ is replaced with $P(t_{post}|i,a_{last})$, the post-relevance given the information need. Furthermore, $(1-\gamma)$ is replaced with $P(t_{pre}|i,a_{next})$, the pre-relevance of $i$ for the next activity.

\subsubsection{Relevance of an information need}
\label{sec:anticipating:model:pia}

The probability of information need relevance given an activity, $P(i|a)$, is based on the relative frequency of the normalized information need in query suggestions for an activity as described in~\S\ref{sec:needs:relevance}, cf. Eq.~\eqref{eq:pia}. 
Additionally, we introduce an extension that takes into account the hierarchy of activities.  Recall that we consider activities at two different levels of granularity: top-level and second-level POI categories (cf.~\S\ref{sec:needs:activities}).
Since we have an order of magnitude more data on the top-level, we expect to get a more reliable estimate for second-level activities ($a^{l2}$) by smoothing them with data from the corresponding top-level activity ($a^{l1}$): 
\begin{equation}
   P_{H}(i|a^{l2}) = \lambda P(i|a^{l2}) + (1-\lambda) P(i|a^{l1}). \nonumber
\end{equation}
Instead of setting $\lambda$ to a fixed value, we use a Dirichlet prior, which sets the amount of smoothing proportional to the number of observations (information needs) we have for the given activity: $\lambda = \beta / (n(a^{l2}) + \beta)$,
where $\beta$ is the total sum of all second-level information needs and $n(a^{l2})$ is the number of information needs for the current (second-level) activity. Using the Dirichlet prior essentially makes this method parameter-free.

\subsubsection{Temporal scope of an information need}
\label{sec:anticipating:model:temp}

We estimated the temporal scope of information needs that belong to the top-level activities with the help of crowdsourcing (\S\ref{sec:temporal}). 
Due to conceptual similarity of activities that are on the same path in the hierarchy, we inherit the temporal scopes of the second-level activities from their top-level parents. This reduces the required crowdsourcing effort by an order of magnitude.

\subsubsection{Transition probabilities}
\label{sec:anticipating:pattern}

In order to anticipate a user's information demands before some next activity, it is necessary to determine which activity it will be. Clearly, some activity sequences are more common than others. For instance, chances are higher that after spending a day working an ordinary person goes home rather than to climb a mountain.  We estimate the likelihood of transition from activity $a$ to activity $b$ (i.e., the dashed arrows in Figure~\ref{fig:model}) by mining a large-scale check-in dataset (cf.~\S\ref{sec:needs:activities:fsdata}). The most frequent transitions between two second-level activities are listed in Table~\ref{tab:top_transitions}.

\begin{table}[t]
    \begin{center}
    \resizebox{\columnwidth}{!}{%
    \begin{tabular}{ c l l r c } 
		\toprule[1.2pt]
        \textbf{Rank} &\textbf{Activity $a_i$} & \textbf{Activity $a_j$} & $n(a_j|a_i)$ & $P(a_j|a_i)$ \\ 
        \midrule
        1. & Train Station	& Train Station & 223246	 	& 0.457 \\         
        2. & Home (private) 	& Home (private) & 106868	& 0.199 \\
        3. & Subway 		& Subway			& 76261		& 0.313 \\         
        4. & Airport 		& Airport		& 70018 		& 0.449 \\         
        5. & Mall 			& Mall 			& 59949		& 0.078 \\                 
        6. & Mall 			& Home (private)	& 46067		& 0.060 \\                 
        7. & Bus Station 	& Bus Station	& 45562 		& 0.188 \\         
        8. & Mall 			& Movie Theater 	& 45360		& 0.059 \\                 
        9. & Office 		& Office 		& 45004		& 0.122 \\                 
        10. & Road	 		& Road	 		& 44752		& 0.167 \\                         
        11. & Food\&Drink Shop & Home (private)	& 38986	& 0.142 \\      
        12. & Mall 			& Food\&Drink Shop 	& 34150	& 0.044 \\    
        13. & Mall 			& Coffee Shop 	& 32572		& 0.043 \\                  
        14. & States\&Municipal. 		& Home (private) 	& 30443	& 0.111 \\                  
        15. & Residential Build.  & Home (private) 	& 27687	& 0.145 \\  
        16. & Home (private)  & Mall 		& 27402		& 0.051 \\
        
        17. & Mall  		& Clothing Store & 25433		& 0.033 \\
        18. & Mall  		& Cafe		 	& 25310		& 0.033\\
        19. & University  	& College Bldg. 	& 24876	& 0.102 \\
        20. & Road  		& Home (private)	 & 24766		& 0.093\\                                                                  
                                                                     
        \bottomrule[1.2pt]
    \end{tabular}
    }
    \tablecaptionshrink	    
    \caption{Most frequent transitions between second-level activities.}    
    \label{tab:top_transitions}    
    \end{center}
\end{table}

Specifically, we define \emph{activity session} as a series of activities performed by a given user, where any two consecutive activities are separated by a maximum of $6$ hours.\footnote{Duplicate check-ins ($1.2\%$ of check-ins), i.e., when a user checks in multiple times at the same place and approximately the same time, were removed before we started to process the sessions.}
We represent activity sequences in a Markov model, allowing the representation of transition probabilities.  The probability of transition from activity $a_i$ to activity $a_j$ is computed using maximum likelihood estimation:
\begin{equation}
\label{eq:trans_prob}
	P(a_j|a_i) = \frac{n(a_i \rightarrow a_j)}{\sum_k n(a_i \rightarrow a_k)},
\end{equation}
where $n(a_i \rightarrow a_j)$ is the number of times activity $a_i$ is followed by activity $a_j$, within the same activity session.

This first order Markov model is a simple, but effective solution.  It yields $80\%$ precision at rank $5$ for top-level and $32\%$ precision at rank $5$ for second-level activities (when trained on $80\%$ and tested on $20\%$ of the check-in data).  We note that more advanced approaches exist for estimating location-to-location transition probabilities, e.g., using higher order Markov models \cite{zhang-2014-lore} or considering additional context \cite{zhang-2015-lta, yuan-2013-tap}.  
Nevertheless, we wish to maintain focus on the core contribution of this work, the anticipation of information needs, which goes beyond next activity prediction.  

\subsection{Experimental setup}
\label{sec:activity:expsetup}

To objective of our last set of experiments is to evaluate how well we can anticipate (i.e., rank) information needs given a past activity.
For this evaluation to be meaningful, it needs to consider what other activity actually followed after in the user's activity session.  That is, we evaluate the relevance of information needs with respect to the transition between two activities.  Since our system is not deployed for real users, testing has to rely on some sort of simulation.  We make this simulation as realistic as possible by taking actual activity sessions from our check-in dataset.

\begin{figure}[ht!]
    \centering
	\includegraphics[width=0.49\textwidth]{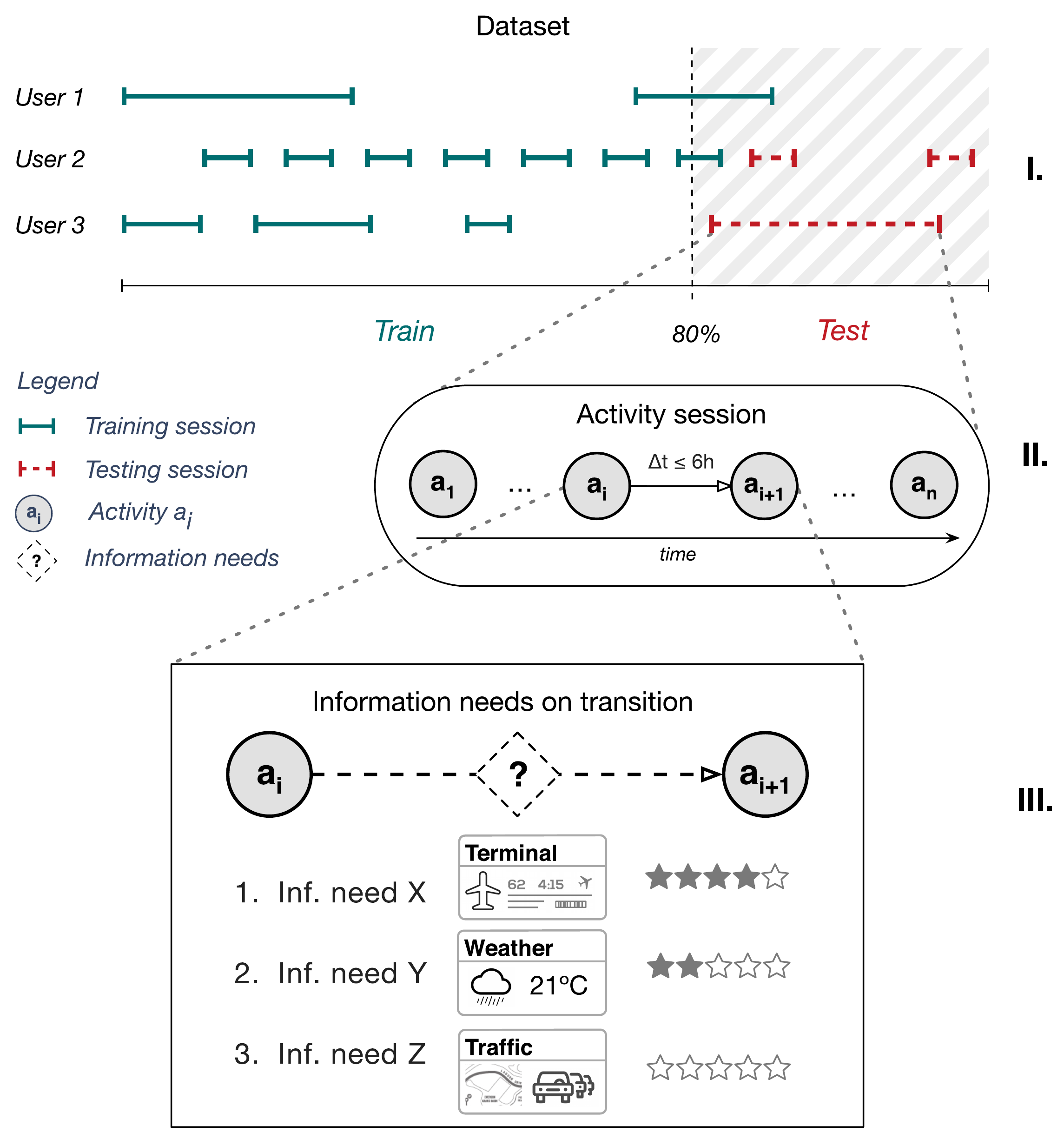}
    \caption{Train/test dataset split (I) with an activity session in detail (II) and an example of assessment interface for evaluating information need usefulness during transition from activity $a_i$ to $a_{i+1}$ (III).}
    \label{fig:evaluation}
    \captionshrink
\end{figure}

Specifically, the check-in data are split into training and testing set, see Figure~\ref{fig:evaluation} part I. The training set consists of the chronologically earlier 80\% sessions and the testing set contains the remaining 20\%.  
The sessions are treated as atomic units so that none of them can be split in half between training and testing. 
The training set is used for establishing the activity-to-activity transition probabilities (\S\ref{sec:anticipating:pattern}).  For each activity session within the test set, we consider transitions for manual evaluation (Figure~\ref{fig:evaluation}, part II), as follows:
(1) for top-level activities, every possible transition between two activities ($9\times 9$) is evaluated;
(2) for second-level activities, due to the large number of possible activity combinations, we take a sample of the $100$ most frequent distinct transitions from the testing fraction of the  check-in dataset.
Crowd judges are tasked with evaluating the usefulness of individual information needs, presented as cards, given the transition between two activities (Figure~\ref{fig:evaluation}, part III).  We collected judgments for the top $10$ information needs from each of the activities in the transition.
See the detailed Algorithm~\ref{alg:eval_alg} below for the process for second-level activities.

\begin{algorithm}
  \caption{Evaluation algorithm (second-level activities)
    \label{alg:eval_alg}}
  \begin{algorithmic}[1]
	\State ${T^{100}} \gets t_1,\dots,t_{100} \in T_{test}$\Comment{ Most frqnt. $2^{nd}$-lev. transit.}
	\State ${n} \gets 10$\Comment{ Number of inf. needs to consider.}
	\State ${results} \gets \emptyset$\Comment{ Store NDCG results per trans.}	
	
	\ForEach {$\text{transition } t=<t_a,t_b> \in T^{100} $}
		\State ${judgments} \gets \emptyset$	
		\State ${needs} \gets top_nneeds(t_a)\cup top_nneeds(t_b)$
		\ForEach {$ n \in needs $} \Comment{ Crowdsourcing assessments.}
			\State $judgments_{t} \gets crowd\_rating(n, t)$ 
		\EndFor
		\State $ranking_t \gets model(t_a)$ \Comment{ Ranking from our model.}
		\State $results \gets ndcg(ranking_t, judgments_t)$	
	\EndFor
	
	\State $avg(results)$\Comment{ Calculate average NDCG value.}	
	
  \end{algorithmic}
\end{algorithm}

\subsection{Evaluation results}
\label{sec:activity:eval}

Since the problem was laid out as a ranking exercise and we make the assumption that most smartphones can certainly display $3$ information cards at once, we use NDCG@3 as our main evaluation measure.  Considering the possibility of scrolling or the use of a larger device, we also report on NDCG@5.  For a baseline comparison, we include a context-agnostic model \textbf{M0}, which always returns the most frequent information needs, regardless of the last activity.  We use a two-tailed paired t-test for significance.

Table~\ref{tab:eval_all} presents the results; corresponding significance testing results (p-values) are reported in Table~\ref{tab:eval_sign}.
All models significantly outperform the naive baseline model (M0) on both hierarchical levels.
Comparing M1, M2, and M3 against each other, we find that M2 outperforms M1; the differences are significant, except NDCG@3 for top-level activities.
As for M3, this more complex model performs significantly worse than M1 on top-level activities and insignificantly better than M1 on second-level activities. M2 always outperforms M3, and significantly so with one exception (second-level NDCG@5).  We suspect that this is due to data sparsity, i.e., we would need better estimations for temporal scope for M3 to work as intended.
Finally, we find that hierarchical smoothing has little (M1 vs. M1-H) to no effect (M2 vs. M2-H and M3 vs. M3-H).  This suggests that the estimations for second-level activities are reliable enough and smoothing does not have clear benefits.

In summary, we conclude that M2 is the best performing model. The value of parameter $\gamma$ is $0.13$ (cf. Eq.~\eqref{eq:gamma}), meaning that the past activity has small, yet measurable influence that should be taken into account when anticipating future information needs.

\begin{table}[t]
    \begin{center}
    \caption{Results for anticipating information needs, second-level activities. The -H suffix indicates the usage of hierarchical smoothing (only for second-level activities). Highest scores are boldfaced.}
    \label{tab:eval_all}
    \tablecaptionshrink
    \begin{tabular}{ @{~}l c c c c c}     
	\toprule[1.2pt]
	\textbf{Model}	& \multicolumn{2}{c}{\textbf{top-level}} & & \multicolumn{2}{c}{\textbf{second-level}} \\
	& \textbf{\small NDCG@3} & \textbf{\small NDCG@5} & & \textbf{\small NDCG@3} & \textbf{\small NDCG@5}  \\
	\midrule
	M0 
	&  0.607 & 0.695 &
	&  0.532 & 0.560  \\ 
	M1 
	&  0.824 & 0.828  &
	&  0.712 & 0.705  \\ 
	M1-H	 
	& & &
	&  0.736 & 0.709  \\ 
	M2 
	&  \textbf{0.852} & \textbf{0.849} &
	&  \textbf{0.765} & \textbf{0.744}  \\ 
	M2-H 
	& & &
	&  \textbf{0.765} & \textbf{0.744}  \\ 
	M3 
	&  0.756 & 0.780  & 
	&  0.735 & 0.741  \\ 
	M3-H 
	& & &
	&  0.735 & 0.740  \\ 	
	\bottomrule[1.2pt]
	\end{tabular}
	\end{center}
    \captionshrink
\end{table}

%
%

\begin{table}[t]
    \begin{center}
    \caption{Significance testing results (p-values).}
    \label{tab:eval_sign}
    \tablecaptionshrink
    \begin{tabular}{ @{~}l c c c c c@{~}}     
	\toprule[1.2pt]
	\textbf{Model}	& \multicolumn{2}{c}{\textbf{top-level}} & & \multicolumn{2}{c}{\textbf{second-level}} \\
	& \textbf{\small NDCG@3} & \textbf{\small NDCG@5} & & \textbf{\small NDCG@3} & \textbf{\small NDCG@5}  \\
	\midrule
	M0 vs. M1   
	& 0.0004 & 0.0068 &
	& 0.0028 & 0.0014 \\
	M0 vs. M2   
	& 0.0002 & 0.0012 &
	& 0.0009 & 0.0004 \\
	M0 vs. M3    
	& 0.0072 & 0.0131 &
	& 0.0073 & 0.0007 \\
	M1 vs. M2  
	& 0.1183  & 0.0264 &
	& 0.0307 & 0.0283 \\
	M1 vs. M3
	& 0.0350 & 0.0541 &
	& 0.8829 & 0.1345 \\
	M2 vs. M3   
	& 0.0021 & 0.0012 &
	& 0.0199 & 0.9553 \\
	\bottomrule[1.2pt]
	\end{tabular}
	\end{center}
    \captionshrink
\end{table}

\subsection{Analysis}
\label{sec:activity:anal}
\begin{figure}[ht!]
    \centering
	\includegraphics[width=0.48\textwidth]{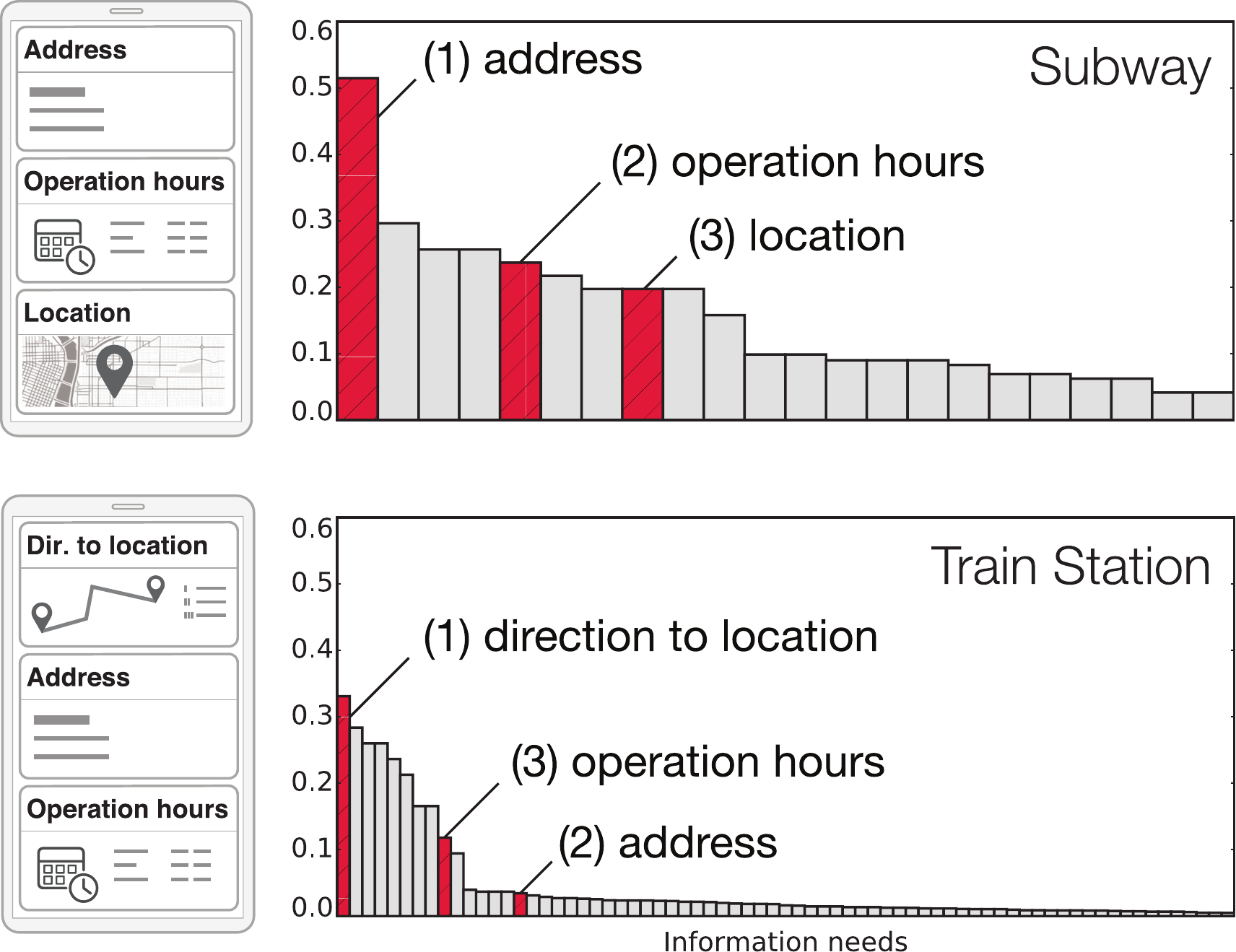}
    \caption{Examples of dashboards and the distribution of the underlying (anticipated) information needs for Subway and Train St.}
    \label{fig:dashboards}
\end{figure}

We take a closer look at two seemingly very similar second-level activities related to transportation. They represent the two extremes in terms of performance using our best model, M2.
Category `Transport/Subway' 
achieve an NDCG@3 score of $0.943$, 
while `Transport/Train Station' only reaches $0.602$.  Figure~\ref{fig:dashboards} shows the corresponding dashboards and the distributions of the information needs to be anticipated according to the ground truth judgments.  
We inspected all individual information needs as predicted by our model and the root cause of the above differences boils down to a single information need: \emph{`address.'} For Subway, 
\emph{`address'} is one of the most important information needs, for all potential transition categories, both according to our model and as judged by assessors. On the other hand, when traveling from a train station, there are $11$ other information needs that are more important than \emph{`address'} according to the ground truth. 
The most likely transition from a train station is another train station (with probability $0.457$).  Even though \emph{`address'} is irrelevant for this transition, overall it still ranks 2$^{nd}$ on the dashboard because of the other transitions that we expect to follow after a train station.
One possible explanation is that when traveling from one train station to another, perhaps covering long distances, a concrete address is not an immediate information need. This is supported by the fact that the more abstract \emph{`city'} is considered important during the `Train Station'$\rightarrow$ `Train Station' transition. When taking a subway, 
it is much more likely that the full address of the next destination is needed.\\\\



\section{Conclusions}
\label{sec:conclusions}

In this paper, we have addressed the problem of identifying, ranking, and anticipating a user's information needs based on her last activity.
Representing activities using Foursquare's POI categories, we have developed a method that gathers and ranks information needs relevant to an activity using a limited amount of query suggestions from a search engine.  Our results have shown that information needs vary significantly across activities.
We have further found in a thorough temporal analysis that information needs are dynamic in nature and tend to change throughout the course of an activity. 
We have combined insights from these experiments to develop multiple predictive models to anticipate and address a user's current information needs in form of information cards.  
In a simulation experiment on historical check-ins combined with human judgments, we have shown that our models have good predictive performance.

In future work we intend to focus on better next-activity prediction by extending the context with time. Previous studies have shown, that mobility patterns are highly predictable \cite{gonzlez-2008-uih}, yet very individual \cite{zhang-2013-igslr}, therefore it would be also interesting to provide personalized results.

\bibliographystyle{abbrvnat}
\renewcommand{\bibsection}{\section{References}}
\setlength{\bibsep}{0pt}
{\raggedright\small
\bibliography{wsdm2017-activity}
}

\end{document}